\documentclass[letterpaper]{article}
\usepackage{nips,times}
\usepackage{amssymb}
\usepackage{graphicx}
\usepackage{listings}
\usepackage{url}
\usepackage{common}
\usepackage[tight,footnotesize]{subfigure}
\newcommand{\g}[1]{``{#1}''}      
\renewcommand{\t}[1]{{\emph{#1}}}  

\begin{document}

\title{Discovering Latent Patterns from the Analysis of User-Curated Movie Lists}

\author{
Derek Greene \\
INSIGHT Centre for Data Analytics\\
University College Dublin, Ireland\\
\texttt{derek.greene@ucd.ie} \\
\And
P\'{a}draig Cunningham \\
INSIGHT Centre for Data Analytics\\
University College Dublin, Ireland\\
\texttt{padraig.cunningham@ucd.ie} \\
}

\maketitle


\begin{abstract}
User content curation is becoming an important source of preference data, as well as providing information regarding the items being curated. One popular approach involves the creation of lists. On Twitter, these lists might contain accounts relevant to a particular topic, whereas on a community site such as the Internet Movie Database (IMDb), this might take the form of lists of movies sharing common characteristics. While list curation involves substantial combined effort on the part of users, researchers have rarely looked at mining the outputs of this kind of crowdsourcing activity. Here we study a large collection of movie lists from IMDb. We apply network analysis methods to a graph that reflects the degree to which pairs of movies are ``co-listed'', that is, assigned to the same lists. This allows us to uncover a more nuanced grouping of movies that goes beyond categorisation schemes based on attributes such as genre or director.
\end{abstract}

\section{Introduction}
\label{sec:intro}

Recently there has been increasing interest in the activity of content curation by users on social media platforms \cite{zhong13loves}. One widely-adopted mechanism for structured content curation is that of \emph{user-curated lists}. From a high-level perspective, the creation of lists can be viewed as a crowdsourced effort to categorise items of interest into user-defined sets, which may or may not have an ordering. The exact semantics of the groups can vary considerably across platforms, and their usage within a specific platform can often also be unclear or inconsistent.  However, by aggregating these lists at a macro-level, there exists the potential to harness this crowdsourced effort to gain a better insight into both the items being curated and the users themselves. 

In the literature, the study of list curation activity has largely focused on the Twitter microblogging platform, which supports the assignment of user accounts to topical lists. Mining information from Twitter lists on a large scale has been used to support tasks such as user recommendation and event detection \cite{greene12recsys}. However, the mass creation of list information is a common activity in many contexts, such as in e-commerce sites where users build ``wish lists'' of their favourite products, or in interest-oriented sites where users create lists of their favourite movies, books, videos, or locations. In some respects, the aggregation of lists is similar to the creation of ``folksonomies'' derived from tagging activity on sites such as \url{delicious.com}, in that it allows a categorisation scheme to naturally emerge without the requirement for exhaustive expert annotation of the data in question.

\begin{figure}[!t]
\centering
\includegraphics[width=0.74\linewidth]{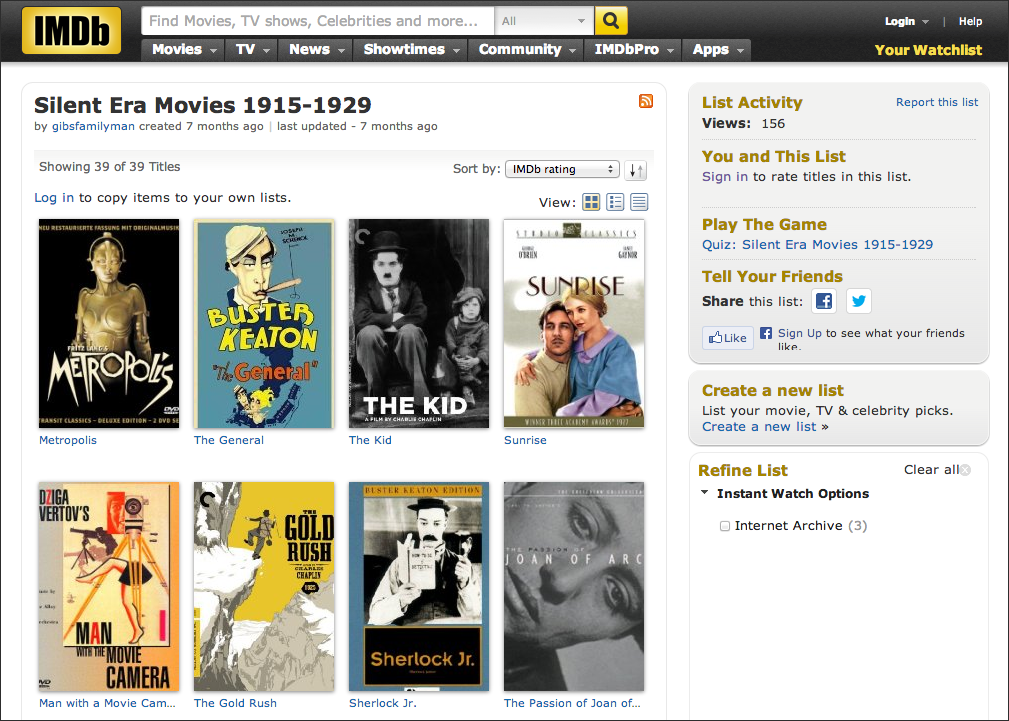}
\vskip -0.3em
\caption{A sample user-curated IMDb list, containing 35 ``Silent Era'' feature films.}
\label{fig:screenshot}
\end{figure}

As a specific instance of list curation activity, we focus on the Internet Movie Database (IMDb)\footnote{\url{http://imdb.com}}, a well-established community interest site that provides access to material relating to films, TV shows, and actors, attracting over 100 million unique users each month \cite{imdbwiki}. The site provides the facility for registered users to create lists containing sets of either movies, games, characters or actors -- see \reffig{fig:screenshot} for an example. Often users create lists that reflect their tastes (\eg ``My favorite Sci-Fi movies of all time''), or that group items according to a particular niche (\eg ``Silent Era Movies 1915-1929''). 
Currently these lists are not summarised on the IMDb website. Also, to the best of our knowledge, IMDb list curation activity and data has not been previously studied in the research community.

In this paper, we perform a large-scale analysis of a collection of $\approx120k$ IMDb lists, curated by $\approx44k$ distinct users, covering over $249k$ movies. We explore the usage patterns of lists by users on the site. Following on from previous work on co-citation analysis in bibliometrics \cite{white81cocite,gmur03invisible}, we show that the use of network analysis methods applied to a normalised \emph{co-listed graph} can be used to discover latent clusters of movies, which go beyond simple attributed-based categorisation schemes. This also provides us with an insight into how users curate content on a site such as IMDb, and suggests future possibilities for using this form of crowdsourced user knowledge in fields such as recommender systems.
\section{Related Work}
\label{sec:related}

Although many social networks and community sites support the curation of lists, relatively little attention has been given to the analysis of this form of data in the literature. In the analysis of Twitter data, researchers have largely focused either on users from the perspective of the content that they produce, or in terms of network representations based on follower relations or retweeting activity. However, preliminary work \cite{kim10lists} suggested that latent relations in Twitter data could be extracted by examining user list data. Wu~\etal \cite{wu11says} showed that user list memberships could be used to organise users into a pre-defined set of categories, such as celebrities and organisations. 
Greene~\etal \cite{greene12recsys} examined the task of expanding Twitter lists to include additional relevant accounts, based on a variety of different user attributes, including information derived from existing list memberships.In related work, the authors proposed an approach to identify topical communities of high-profile users on Twitter, by aggregating their list memberships and applying overlapping community finding \cite{greene12ecml}. 
Garc\'{i}a-Silva \cite{garcia12lists} described approaches for extracting semantic relations from user lists, by constructing associations between co-occurring keywords taken from list names. 

While IMDb has previously been considered as a source of data by researchers, it has been primarily used to provide benchmark or auxiliary data for work in recommender systems and visualisation.  Basu~\etal \cite{basu98rec} described a recommendation approach combining user ratings information with metadata attributes retrieved from IMDb, such as genres, keywords, and titles.
Herr~\etal \cite{herr07movies} performed a high-level analysis of the bipartite movie-actor network,  for a set of $\approx 428k$ movies from IMDb as of 2005. The authors focused on visualising the corresponding \emph{co-actor} graph, where a pair of actors are connected by an edge if they appeared in the same movie, and edges are weighted by the number of movies in which they appeared together.
Other work has looked at the potential of user-generated content on IMDb to provide an insight into the attitudes of the site's users towards contemporary cultural issues \cite{dodds06bond}. Despite the use of IMDb in many contexts, we are unaware of work that has considered the use of the site's large repository of lists.
\section{Movie Graph Representation}
\label{sec:methods}

In this section, we discuss the collection of a dataset of user-curated movie lists, and then we describe the creation of a suitable aggregated graph to support the application of network analysis techniques to explore IMDb list data. A pre-processed version of the dataset is available online\footnote{\url{http://mlg.ucd.ie/movielists}}. 

\subsection{Data Collection}
\label{sec:data}

We collected data from IMDb via the ``Newest Lists'' and ``Tagged Lists'' pages, together with lists  indexed by Google. Collection was restricted to lists covering items such as feature films, documentaries, and TV shows/episodes. For readability purposes, from now on we use the term \emph{movies} to refer to these types of items. Items such as video games, actors or characters are not included in this study. At the end of the  process, we had retrieved 121,752 relevant user lists created by 44,097 users, containing 249,261 movies, yielding a total of 7,434,220 list membership entries. The vast majority of users (96.7\%) were responsible for creating $\leq 10$ lists. The mean number of lists per movie was 29.8, with 95.6\% of these assigned to $\leq 100$ lists, while a small number of movies (1,486) were assigned to $\geq 1000$ lists. The most frequently-listed film was \t{The Dark Knight}, which was assigned to 8,591 lists in the dataset. 

For each movie, we also retrieved eight metadata attributes via the IMDb API. Seven of these are categorical:
\begin{itemize}
\item \emph{Type:} The item's basic type -- one of the following: TV movie, feature film, short film, TV episode, TV series, mini-series, video, documentary, TV special.
\item \emph{Decade:} We retrieved the year of the movie's initial release. In the case of a TV series, this corresponds to the airing of its first episode. Individual years were then binned by decade.
\item \emph{Genres:} Each movie can be assigned one or more class of genre (\eg comedy, western etc). IMDb currently has 27 different genres.
\item \emph{Countries:} Set of countries associated with a movie. Our dataset contains movies from 222 different countries.
\item \emph{Languages:} Set of audio languages for the movie. Our dataset contains movies from 291 different languages.
\item \emph{Directors:} Set of directors associated with a movie. Our full collection of movies yielded 1,9162,470 directors.
\item \emph{Actors:} Set of actors associated with a movie. Our full collection of movies has 238,100 associated actors.
\end{itemize}
We also retrieved each movie's \emph{IMDb rating}, a numeric score $\in [1,10]$ that is calculated from a weighted aggregation of user-submitted ratings. A higher value indicates a higher level of popularity among IMDb users. 
\subsection{Data Characterisation}

\begin{figure}[!t]
\centering
\includegraphics[width=0.9\linewidth]{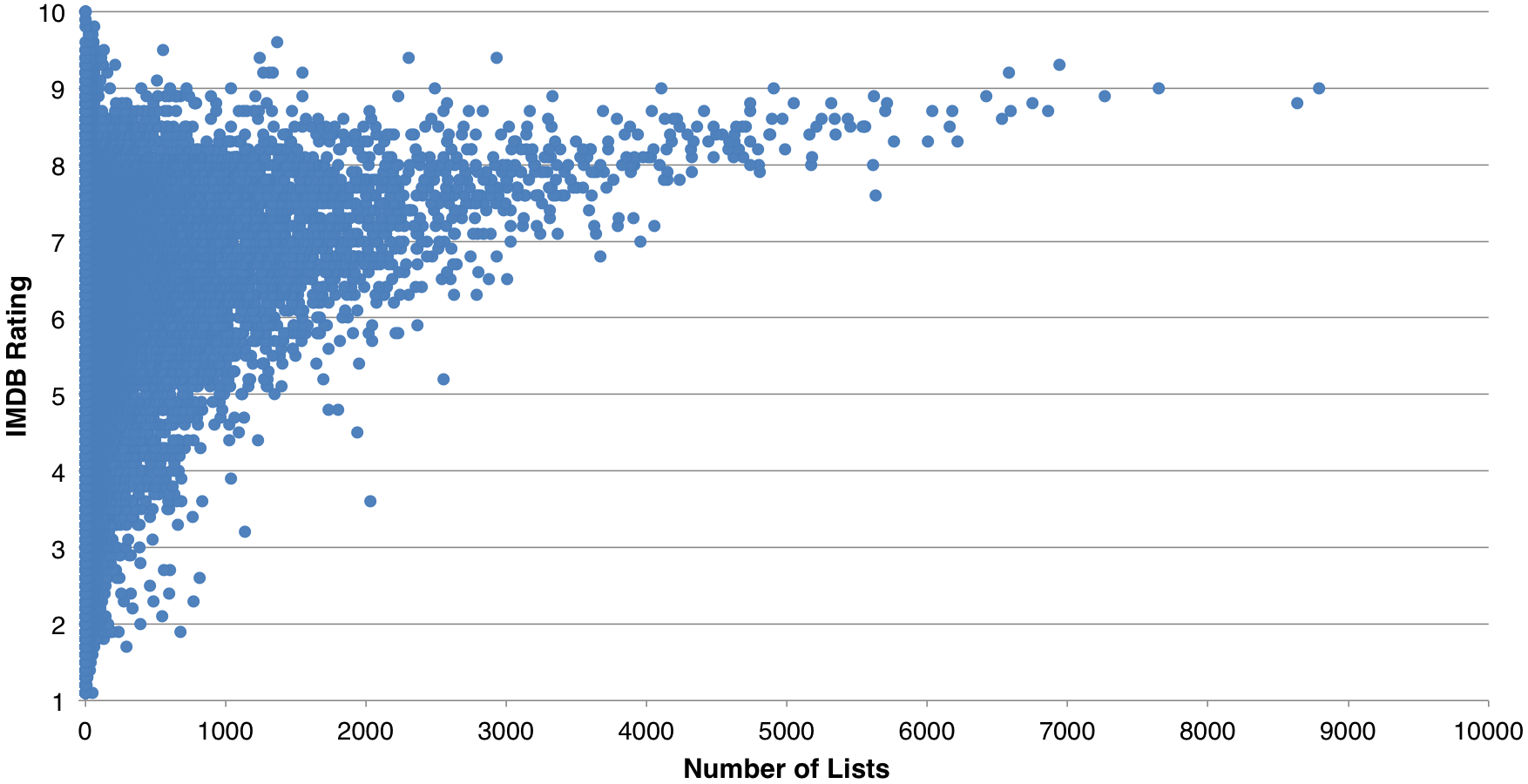}
\caption{Relationship between IMDb moving rating versus number of list assignments per movie, for all 249k movies.}
\label{fig:plot1}
\end{figure}

Prior to analysing the co-listed relations in the data, we explored basic summary statistics of the full dataset to get a sense for list usage in relation to movies. The majority of movies (93\%) were assigned to $\leq 50$ lists, with a very small proportion of frequently-listed movies (0.6\%) assigned to $> 1000$ lists.
\reffig{fig:plot1} shows the distribution of list assignments in the dataset, plotted against corresponding IMDb movie rating scores. The plot indicates a strong correlation between the two, where number of list assignments is generally indicative of rating scores, particularly in the case of highly-rated movies. However, there are a small number of notable outliers, corresponding to movies assigned to over 1,000 lists, but with low user rating scores (\eg \t{Cat Woman}, \t{Batman \& Robin}, \t{The Twilight Saga: New Moon}). \reftab{tab:toplist} summarises the movies that were most frequently assigned to lists, which correspond to points on the extreme right-hand side of the plot in \reffig{fig:plot1}.

\begin{figure}[!t]
\centering
\includegraphics[width=0.9\linewidth]{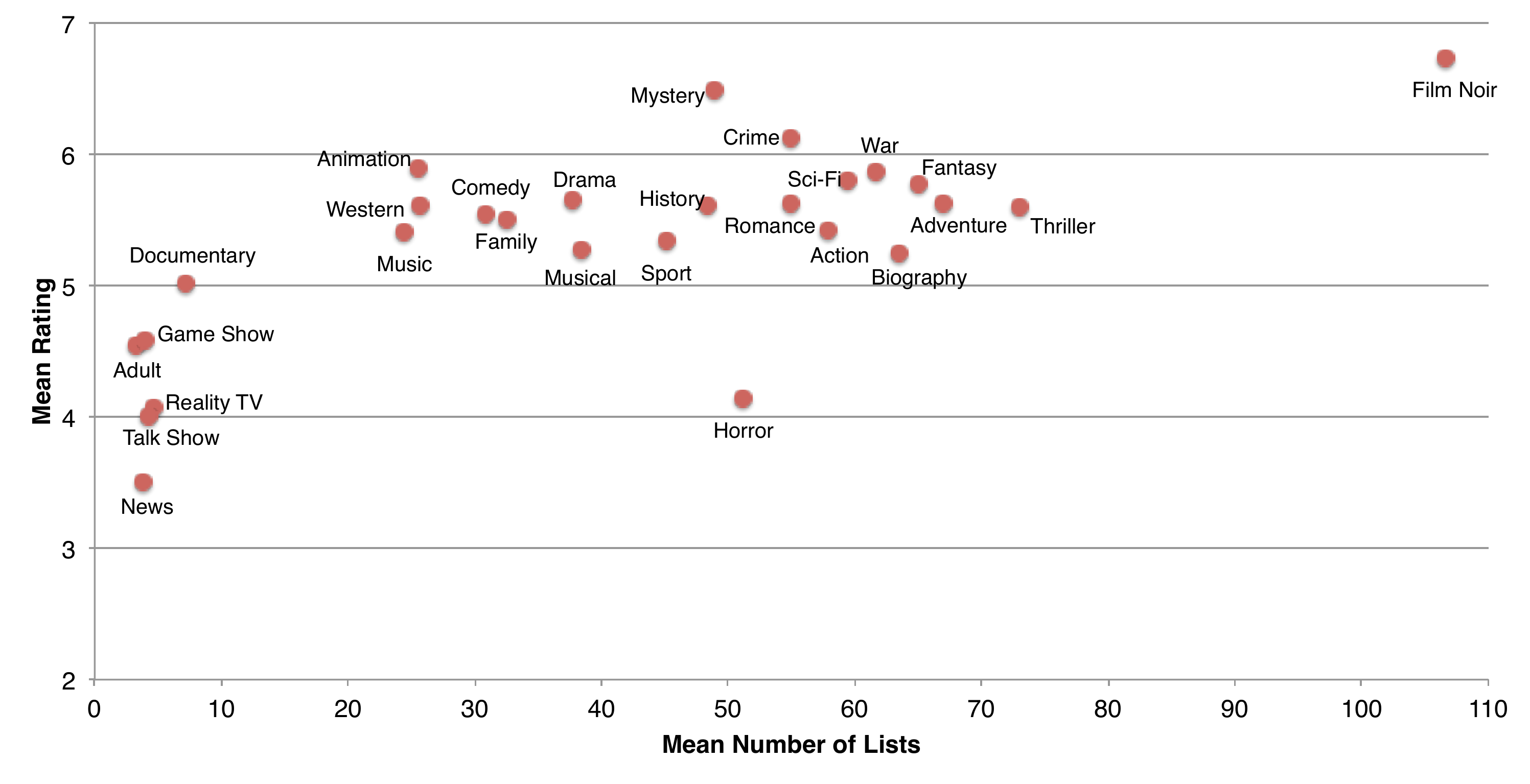}
\caption{Relationship between IMDb moving rating versus number of list assignments per movie, as averaged by movie genre.}
\label{fig:plot2}
\end{figure}

To investigate whether the correlation between list assignment counts and IMDb rating was present across different genres, we computed mean counts and scores for all movies assigned to each IMDb genre. \reffig{fig:plot2} shows a plot of the two mean values as aggregated over movie genre. In general we see the same strong correlation between the two measures. However, the genre ``horror'' is a clear outlier, where the mean rating of movies is considerably lower than expected. On inspection, this is largely due to the presence of low rated ``B-movie'' feature films in this category. We also observe that the small niche genre ``film noir'', largely consisting of crime and drama feature films from 1930s-50s, has far higher list assignments and higher average movie ratings than other IMDb genres.

\begin{table}

    \center
    \begin{tabular}{|c|l|c|c|}
    \hline
    \emph{\#} & \emph{Movie Title}                                             & \;\emph{Lists}\; & \emph{Rating} \\\hline
    1        & The Dark Knight                                   & 8791   & 9.0    \\
    2        & Inception                                         & 8639   & 8.8    \\
    3        & Pulp Fiction                                      & 7652   & 9.0    \\
    4        & Fight Club                                        & 7266   & 8.9    \\
    5        & The Shawshank Redemption                          & 6944   & 9.3    \\
    6        & The Matrix                                        & 6861   & 8.7    \\
    7        & The Lord of the Rings: The Fellowship of the Ring\;\;\;\; & 6750   & 8.8    \\
    8        & Forrest Gump                                      & 6596   & 8.7    \\
    9        & The Godfather                                     & 6583   & 9.2    \\
    10       & The Dark Knight Rises                             & 6532   & 8.6    \\
    11       & The Lord of the Rings: The Return of the King     & 6422   & 8.9    \\
    12       & The Avengers                                      & 6217   & 8.3    \\
    13       & Se7en                                             & 6178   & 8.7    \\
    14       & The Shining                                       & 6160   & 8.5    \\
    15       & The Silence of the Lambs                          & 6039   & 8.7    \\
    16       & Inglourious Basterds                              & 6006   & 8.3    \\
    17       & Batman Begins                                     & 5760   & 8.3    \\
    18       & Star Wars                                         & 5713   & 8.8    \\
    19       & The Lord of the Rings: The Two Towers             & 5701   & 8.7    \\
    20       & Titanic                                           & 5635   & 7.6    \\\hline
    \end{tabular}
    \caption{Ranking of top 20 movies most frequently assigned to IMDb lists, together with their corresponding IMDb rating scores.}
\label{tab:toplist}
\end{table}

\subsection{Co-Listed Graph}
\label{sec:colist}

\begin{figure}[!h]
\centering
\includegraphics[width=0.3\linewidth]{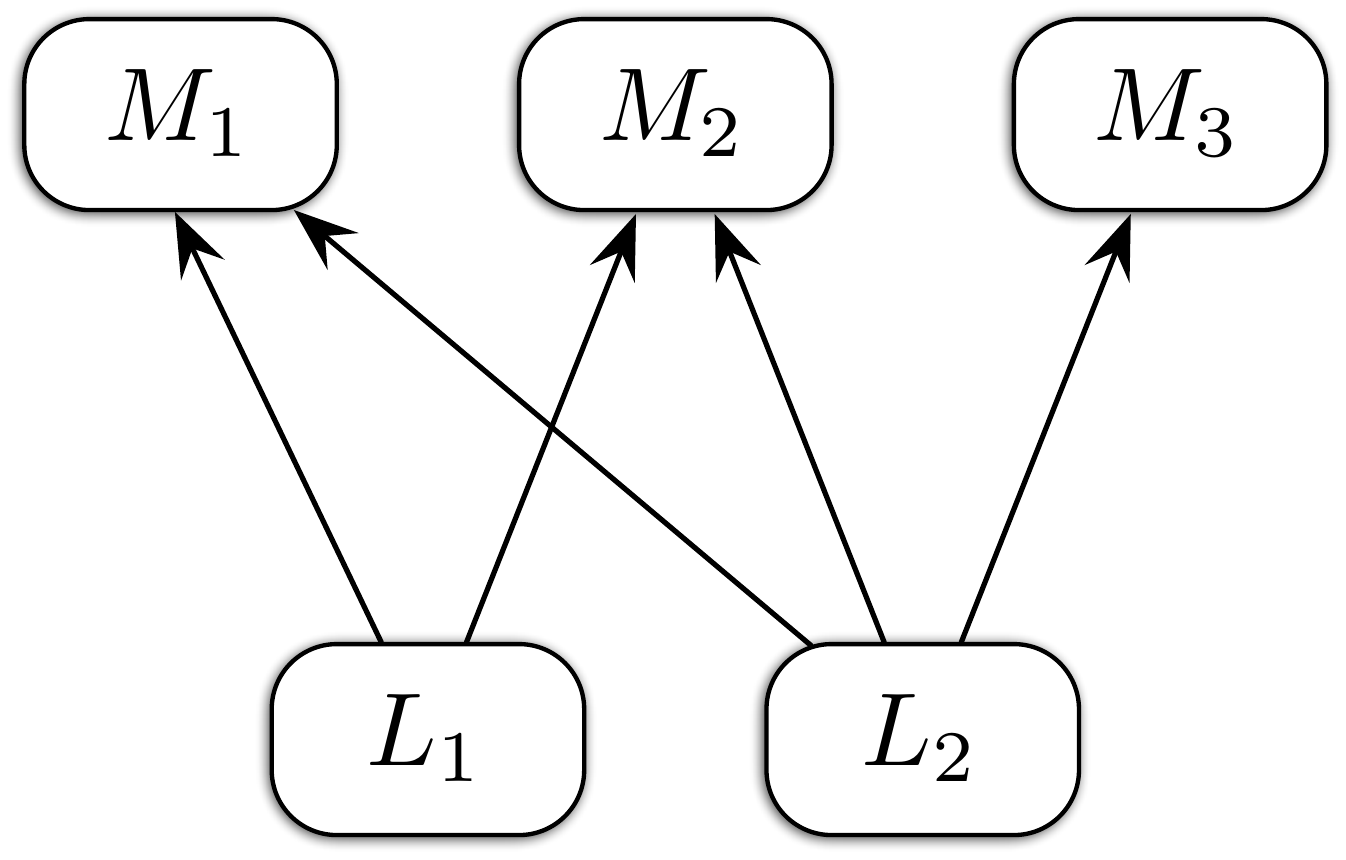}
\caption{Co-listed information can be effective in revealing relationships between movies. In this example, the fact that movies $M_1$ and $M_2$ are assigned to lists $L_1$ and $L_2$ is indicative of a relationship between them. (Note that an arrow from $L_i$ to $M_j$ indicates that the list $L_i$ contains movie $M_j$.)}
\label{fig:colist}
\end{figure}

We now focus on the construction of a graph representation that summarises the most salient information from their IMDb list assignments. 
Following previous work on co-citation analysis \cite{white81cocite}, we propose an approach for extracting underlying associations between curated items that are \emph{co-listed}, in our case movies. The approach is illustrated in Figure \ref{fig:colist}. The fact that the movies $M_1$ and $M_2$ are co-assigned to both lists $L_1$ and $L_2$ indicates a strong implicit relationship between them. In contrast, co-list analysis suggests a weaker relationship between the pairs $(M_{1},M{3})$ and $(M_{2},M_{3})$, which are only assigned together to the same list once. 

In general, we suggest that co-list analysis has the potential to reveal indirect associations between movies that may not be immediately evident from simply comparing their descriptions or individual metadata attributes.

To filter lists without a coherent theme (\eg ``All movies I have watched''), we only considered lists containing $\geq 5$ and $\leq 100$ movies. We also focused on movies with a baseline level of popularity among IMDb users -- those assigned to at least five lists. From the remaining 40,285 movies, we constructed an undirected \emph{co-listed} graph as follows. Each node represents a movie, and an edge exists between two nodes if the corresponding movies have been assigned to the same unfiltered list at least once. Edges weights correspond to the number of lists that the pair of movies share. 

\subsection{Normalised Co-Listed Graph}
\begin{figure}[!t]
\center
\subfigure[]{\includegraphics[width=0.40\linewidth]{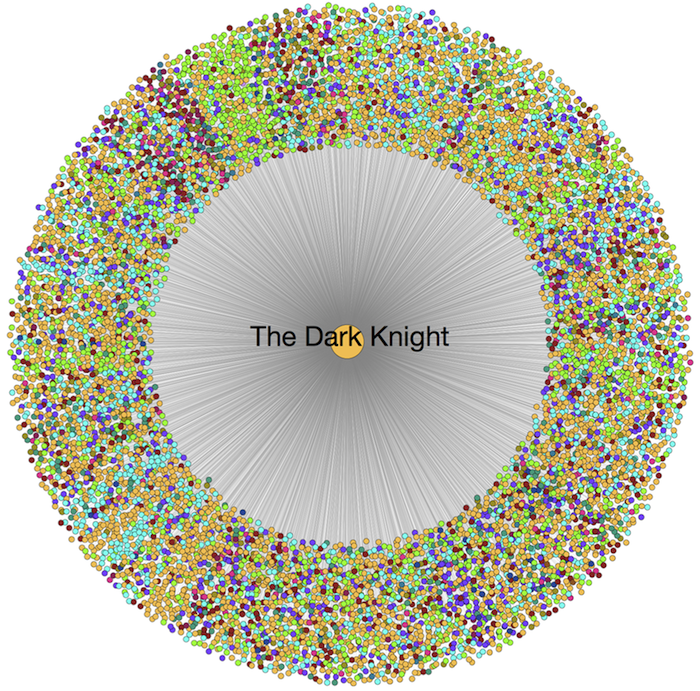}}
\hskip 0.8em
\subfigure[]{\includegraphics[width=0.46\linewidth]{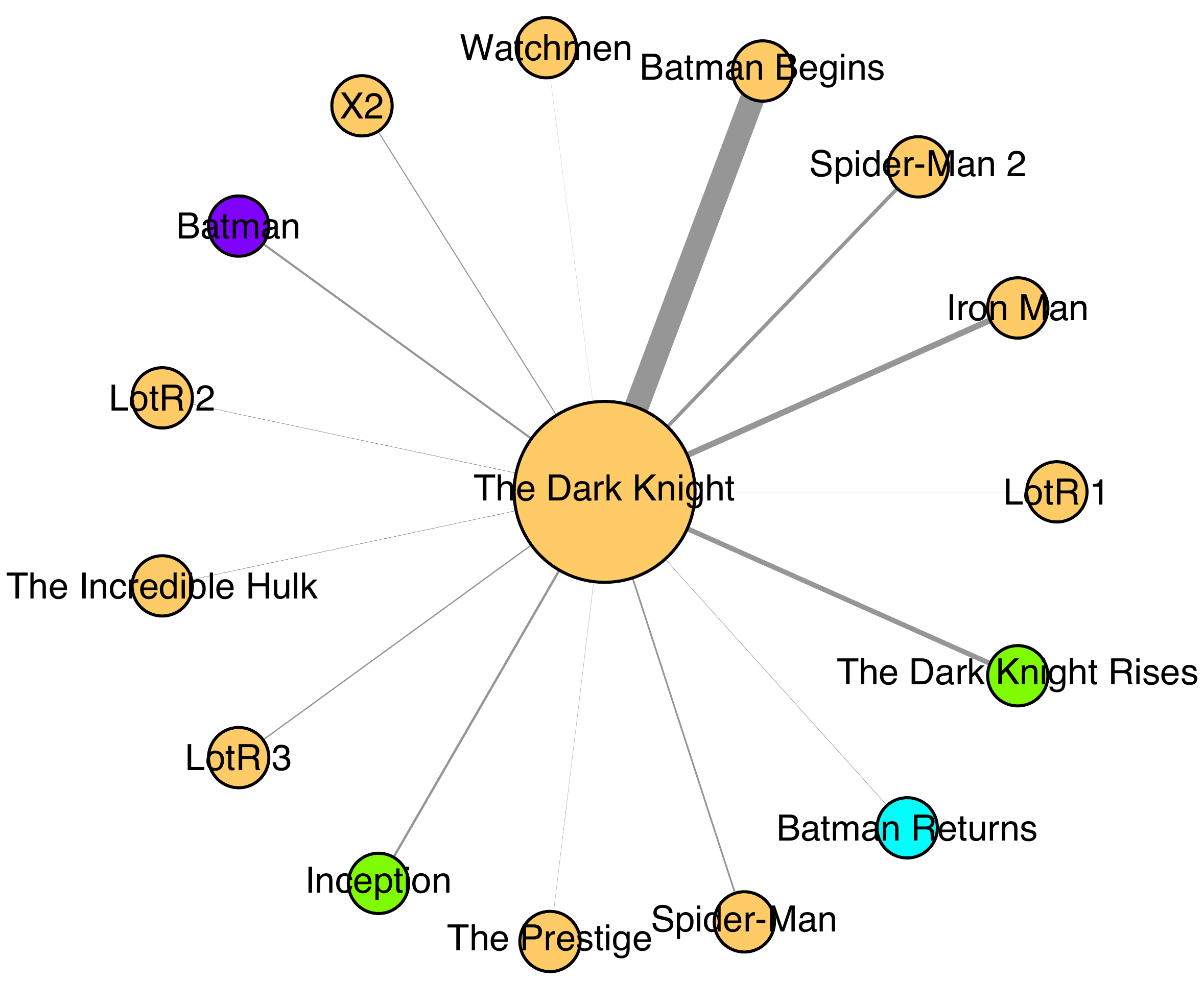}}
\caption{A visualisation of ego networks for a frequently-listed movie, \t{The Dark Knight} in (a) the original unfiltered co-listed graph, (b) the thresholded normalised co-listed graph. Nodes are coloured according to the decade in which the corresponding movie was released.}
\label{fig:hubs}
\end{figure}

In co-citation analysis, hub nodes can often dominate the graph, as they have connections to a wide variety of weakly-related nodes \cite{gmur03invisible}. This can be problematic when attempting to identify latent clusters within the graph. We examined the movie co-listed graph and observed similar behaviour. An example is shown in \reffig{fig:hubs}, focusing on \t{The Dark Knight}, which has the highest number of list assignments in our overall dataset. In \reffig{fig:hubs}(a) we see the ego network for this film in the co-listed graph, with edges to 9,894 other movies. Although the film was released in 2008, it shares lists with movies from every decade as far back as the 1900s. This suggests a lack of homogeneity in the ego network, which might be explained by the film frequently appearing on ``favourite movie lists'' (as of August 2013, this film has an IMDb rating of 9.0, and is 6th in the IMDb Top 250 ranking). 
While we could filter raw co-listed counts, the scale of these values will vary significantly from one movie to another, with no upper bound, making threshold selection unclear.

Therefore, to normalise the edge weights, we employ a technique analogous to the \emph{CoCit-Score}, proposed by Gm\"{u}r \cite{gmur03invisible}, which has been shown to be a particularly effective choice for clustering co-citation data. In the case of co-listed data, the weight of an edge between a pair of movies $M_{A}$ and $M_{B}$ is defined as:
\[
W(M_{A},M_{B}) = \frac{ (|\aset{L}_A \cap \aset{L}_B|)^2 }{ min(|\aset{L}_A|,|\aset{L}_B|) \times mean(|\aset{L}_A|,|\aset{L}_B|) }
\]
where $\aset{L}_A$ is the set of lists containing movie $M_{A}$ and $\aset{L}_B$ is the set of lists containing movie $M_{B}$. The normalisation with respect to mean and minimum list set size results in a weight $\in [0,1]$, where a higher value indicates a greater level of affinity between two movies, in terms of their list assignments.

From our original dataset, the normalisation process yielded a dense weighted graph with 1,659,353 edges. To increase the sparsity of the graph, we subsequently applied an edge threshold of $0.1$. This filtering yielded a final normalised co-listed graph $G$ with 27,664 nodes connected via 162,888 edges. One large connected component in this graph contained $83\%$ of the nodes, with 1,187 far smaller components also present. 

Reverting to our example in \reffig{fig:hubs}, we examine \t{The Dark Knight} as represented in the normalised co-listed graph with threshold $0.1$. We see that the ego network (\reffig{fig:hubs}(b)) now only contains 15 nodes, the majority of which correspond to movies from the 2000s. On closer inspection, we see that edges exist between \t{The Dark Knight} and other Batman films spanning multiple decades, with a particularly strong connection to the prequel film \t{Batman Begins}.

\section {Cluster Analysis}
\label{sec:eval}

\subsection{Methodology}
\label{sec:cluster}

To identify overlapping clusters of movies within the normalised co-listed graph, we apply the popular OSLOM community finding algorithm. However, it has been observed that OSLOM can produce unstable results \cite{lanc12consensus}.  
Following previous work on unsupervised ensembles \cite{lanc12consensus}, we addressed this issue as follows. We constructed a symmetric \emph{consensus matrix} $\m{M}$, with rows and columns corresponding to the movies in the normalised graph. To generate an ensemble of $r$ \emph{base clusterings}, we applied  OSLOM with the default parameter values using a different random seed for each run, to a subset of nodes. A subset of $80\%$ of nodes was randomly sampled for inclusion in each run, and OSLOM was applied to the subgraph induced by this set. 

After generating each base clustering, for every pair of nodes in the graph $(M_{A},M_{B})$, we computed the Jaccard similarity  between the sets of cluster labels assigned to those nodes by OSLOM. If the pair are not both co-assigned to any cluster, the score is 0. If the pair are present in all clusters together, the score is 1. However, unlike the binary approach of \cite{lanc12consensus}, if the pair are present in some but not all clusters together, the Jaccard score will reflect this. After computing scores for all pairs, we incremented the corresponding matrix entries in $\m{M}$. 
We repeated this process to produce $r=100$ base clusterings. The resulting matrix $\m{M}$ was then normalised by $1/r$ so that the entries were $\in [0,1]$. 

To find \emph{consensus clusters}, we followed a similar approach to that used by \cite{lanc12consensus}, although we do not require the use of a user-defined threshold to apply to the consensus matrix.  Rather, we view $\m{M}$ as an affinity matrix, and applied a final run of OSLOM to the weighted graph constructed from $\m{M}$. Singletons and pairs were discarded, leaving 918 consensus clusters ranging in size $[3,143]$, covering 20,965 movies. Of these, 1,703 movies $(8.1\%)$ were assigned to more than one cluster.

\subsection{Macro-Level Validation}
\label{sec:val}

To determine whether the consensus clusters are coherent in terms of their content, we apply a validation process that examines the \emph{enrichment} of each cluster relative to the categorical metadata attributes associated with movies in the cluster. That is, for a given attribute value, we look at the fraction of movies in a cluster that share that same value. We make use of the categorical metadata attributes described  in \refsec{sec:data}.

We observed that certain attribute values are highly-enriched within the dataset as a whole, relative to other values. For instance, the complete dataset contains $\approx113k$ movies in English, while there are only 66 movies in Kurdish. To reduce the possibility of achieving high enrichment scores by chance, we employ the widely-used adjustment technique introduced by Hubert \& Arabie \cite{hubert85compare}: 
\[
\textrm{Corrected\_Enrichment} = \frac{\textrm{Enrichment} - \textrm{Expected\_Enrichment}}{1 - \textrm{Expected\_Enrichment}}
\]

\begin{figure}[!h]
\centering
\includegraphics[width=0.8\linewidth]{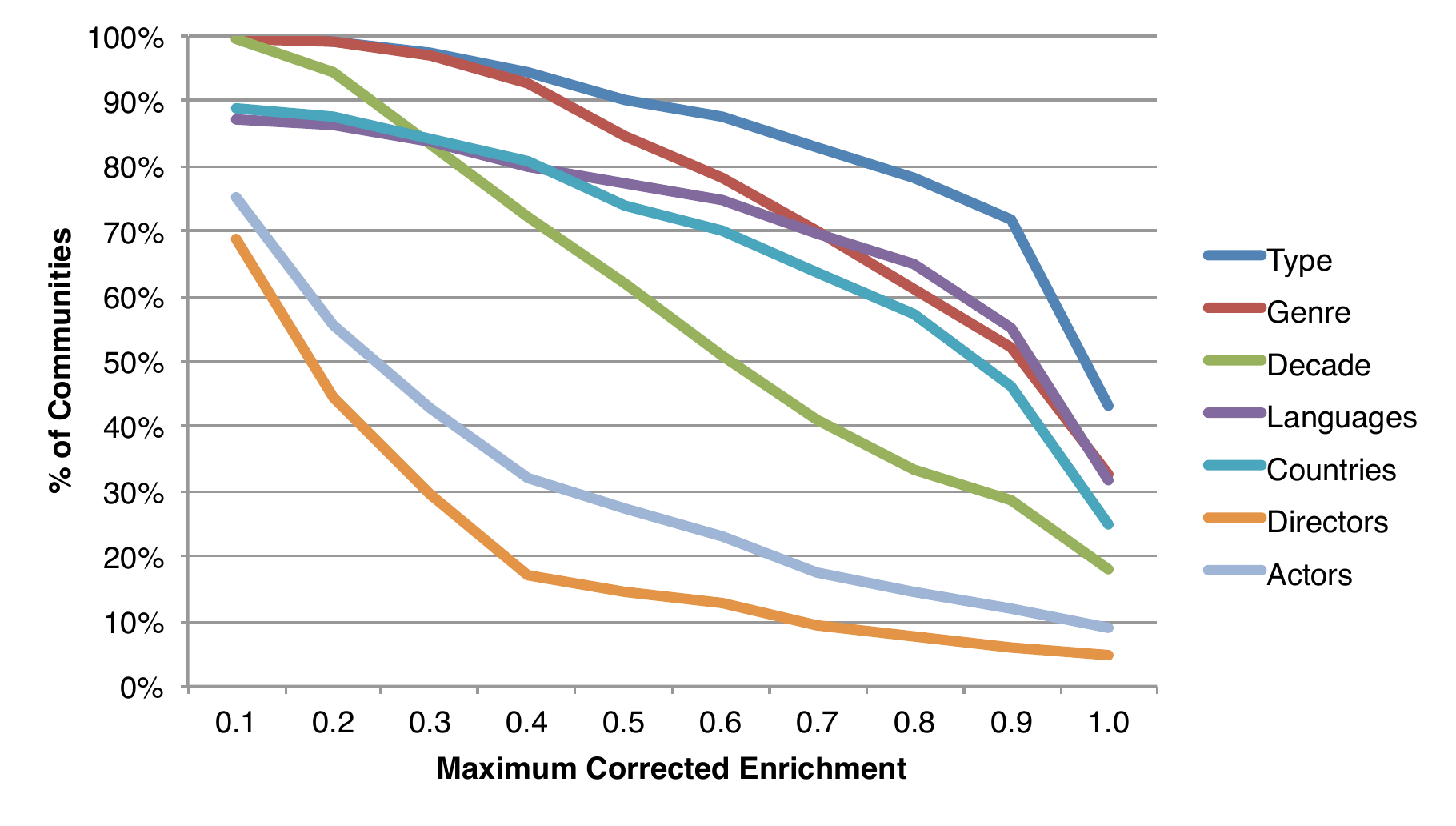}
\vskip -1em
\caption{Maximum enrichment scores $[0.1,1.0]$ achieved by different percentages of clusters, for each of seven categorical movie metadata attributes.}
\label{fig:enrich1}
\end{figure}

\begin{figure}[!h]
\centering
\includegraphics[width=0.8\linewidth]{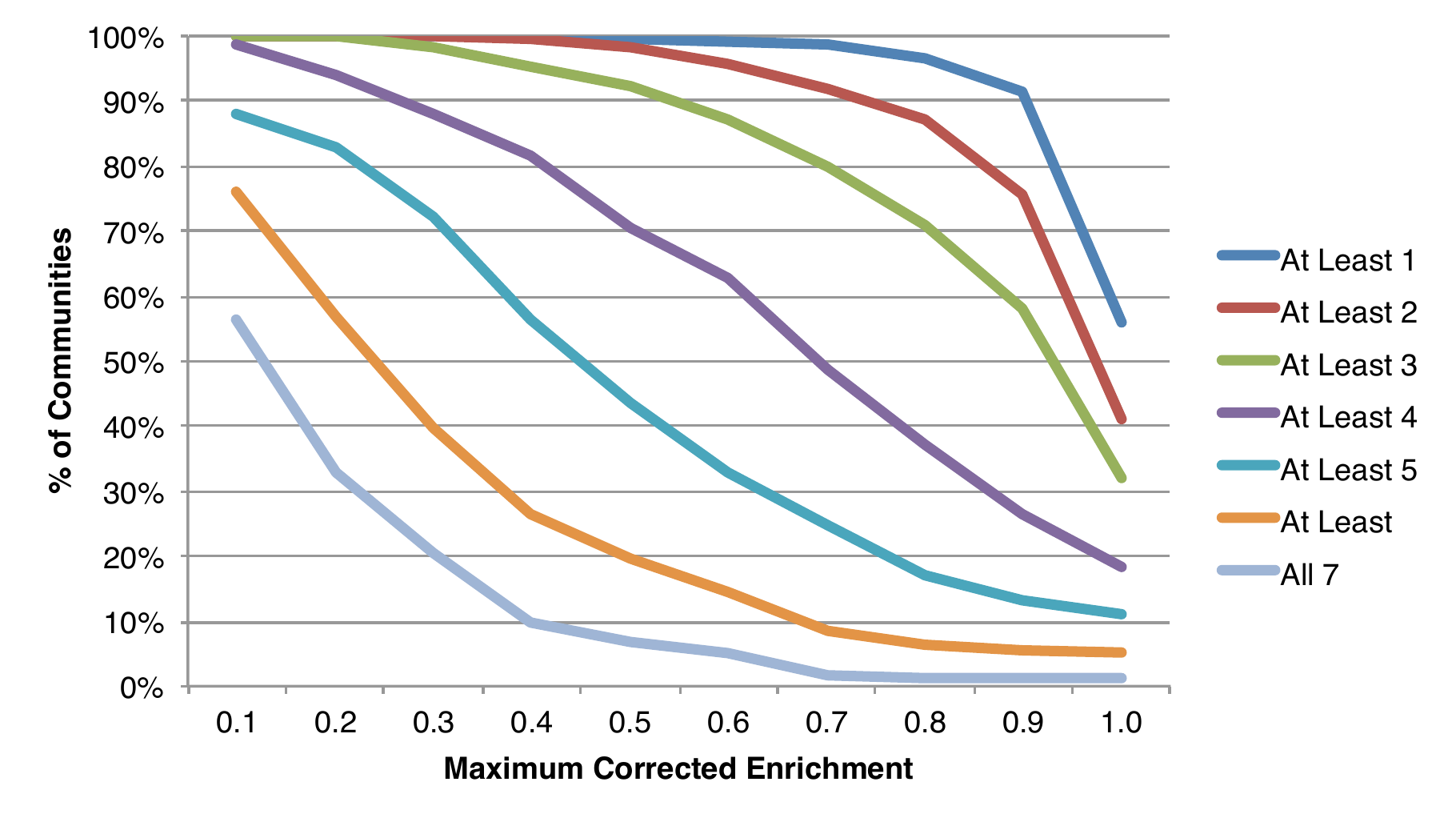}
\vskip -1em
\caption{Maximum enrichment scores $[0.1,1.0]$ achieved by different percentages of clusters, for increasing numbers of combined metadata attributes.}
\label{fig:enrich2}
\end{figure}

For each cluster we calculated the \emph{maximum corrected enrichment} level for any value for a given metadata attribute. For instance, for a cluster that is enriched for several genres (\eg comedy$=0.8$, action=$0.4$, adventure=$0.3$), the maximum enrichment value for the genre attribute is $0.8$. We then computed the percentage of clusters achieving at least a certain level of enrichment $[0.1,1.0]$ for each categorical attribute. 

A summary of results is provided in \reffig{fig:enrich1}. We observed that many clusters achieve high levels of enrichment ($\geq 0.8$) for a number of the attributes. As we might expect, given the very high numbers of actors and directors presents across all movies in the graph, the enrichment levels for these two attributes are comparatively low (14\% and 8\% respectively for enrichment threshold $\geq 0.8$). However, for attributes ``type'', ``genres'', ``languages'', and ``countries'', we found that many clusters reach this level of enrichment, ranging from 57\% to 78\% of all clusters. For example, we identified a cluster of size 30 almost entirely containing comedy feature films from the United Kingdom directed by Gerald Thomas and featuring Kenneth Williams (\ie high maximum corrected enrichment in genre, type, country, director, and actor attributes).

One of our goals is to identify emergent patterns that extends beyond groups of movies that are simply enriched for a single attribute. To explore this at a macro-level, we looked for clusters that were enriched for at least $X$ metadata attribute values. This allowed us to measure the degree to which clusters found on the co-listed graph can capture interesting groupings that reflect highly-specific combinations of metadata attribute values. The summary of these results provided in \reffig{fig:enrich2} indicates that clusters with multiple highly-enriched attributes do exist. We see that 87\% of all clusters are enriched at level $0.8$ for at least two metadata attributes, with 159 clusters (17\%) enriched for combinations of five or more attributes.

\subsection{Micro-Level Analysis}
\label{sec:micro}

To explore individual clusters, we firstly summarise each cluster based on the corrected enrichment scores for metadata attributes. To automatically produce enrichment-based summaries, for each cluster we identify metadata feature values achieving $\geq 0.8$ enrichment. For instance, a cluster of feature films that is enriched for a combination of a specific genre, language, country, and decade might have a summary such as ``Drama feature films in Japanese from Japan from the 1950s''. Similarly, a cluster of films enriched for a specific actor and director might have a summary such as ``Feature films directed by George Miller featuring Mel Gibson''. These automatically-generated summaries helped us to understand the clusters generated on our dataset. Based on this exploration, we now discuss three notable examples of structures discovered via co-listed analysis.

\vskip 1em
\subsubsection{Case Study -- UK TV Series}

\begin{figure}[!h]
\centering
\includegraphics[width=0.72\linewidth]{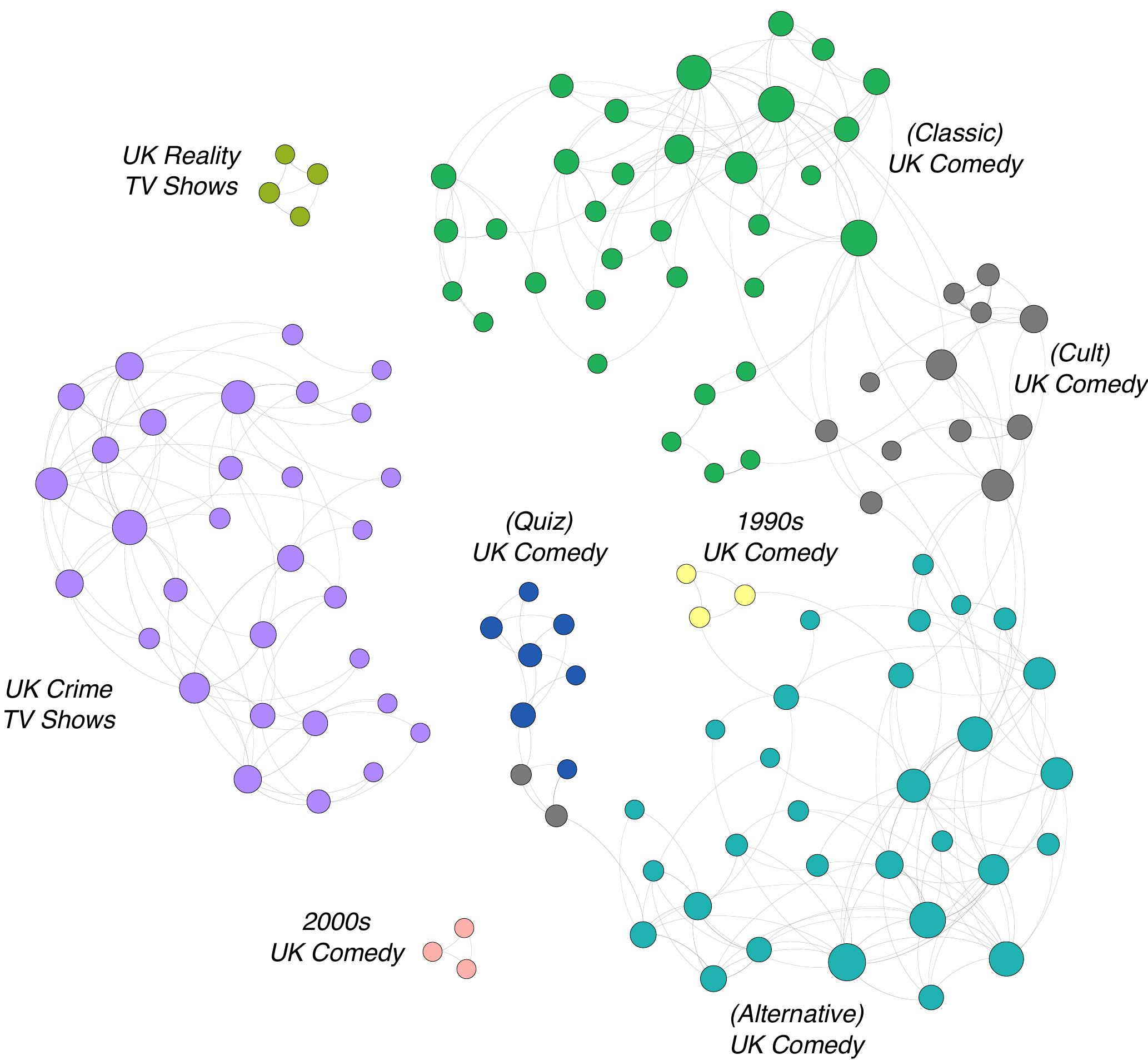}
\caption{A subgraph of the normalised co-listed graph, induced by a set of 7 clusters enriched with TV series originating from the United Kingdom. Nodes are coloured by cluster and scaled by degree. Nodes coloured in grey are assigned to multiple clusters. Cluster annotations were derived from both high enrichment $(\geq 0.8)$ in metadata feature values, and by the movies assigned to the clusters. }
\label{fig:uktv}
\end{figure}

For our initial case study, we examine a set of seven clusters enriched with type \g{TV series} and country \g{United Kingdom}. \reffig{fig:uktv} shows a force-directed visualisation of these clusters. On the left-hand side of the graph, we see two distinct clusters covering crime dramas and reality TV respectively. On the right-hand side, we see a single component consisting of a number of overlapping clusters containing TV shows that all share the genre \g{comedy}. The large cluster at the top corresponds to `classic' TV comedy series from the 1950s until the 1990s, such as \t{Steptoe and Son} and \t{Porridge}. The large cluster at the bottom includes TV series from the 1990s and 2000s which could be described as `alternative British comedy', such as \t{Brass Eye} and \t{Spaced}. The region where these two clusters overlap contains popular series spanning a number of decades which could be described as `cult British comedy', including comedies such as \t{Monty Python's Flying Circus}, \t{Fawlty Towers}, and \t{Father Ted}. 

On the left fringe of the right-hand component we see a small cluster containing series that fall into the narrow niche of comedic and satirical TV quiz shows, which have been popular among UK audiences in recent years (\eg \t{QI}, \t{Have I Got News For You}). In general, the list assignments that result in the right-hand component reflect the differing tastes in comedy among various demographics and different definitions of humour, in a way that is more nuanced than the catch-all \g{comedy} IMDb genre.

\begin{figure}[!t]
\centering
\vskip 0.5em
\includegraphics[width=0.95\linewidth]{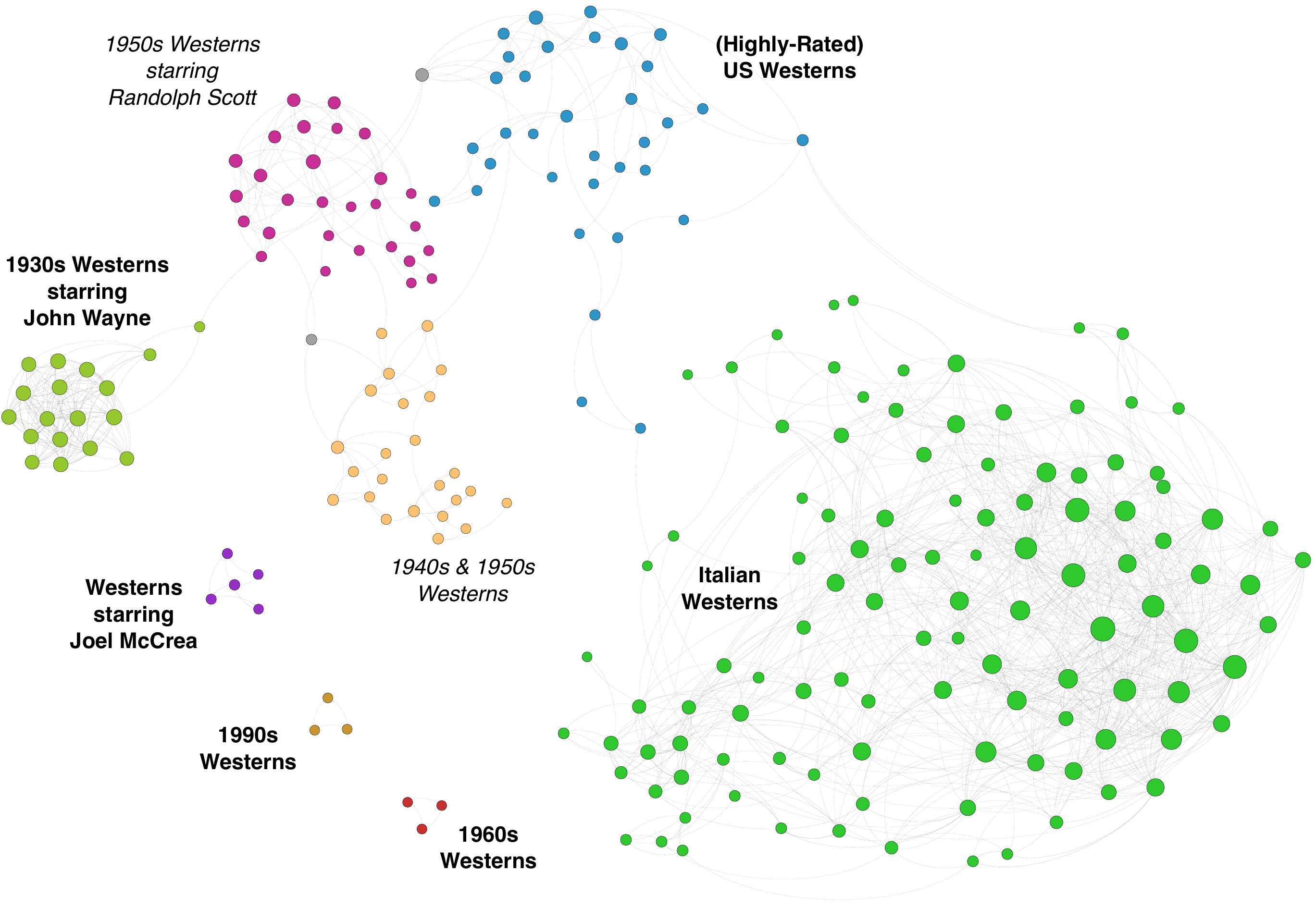}
\caption{A subgraph of the normalised co-listed graph, induced by a set of 8 clusters enriched with Western feature films. Nodes are coloured by cluster and scaled by degree. Nodes coloured in grey are assigned to multiple clusters. Cluster annotations are shown -- those in bold were derived from high enrichment $(\geq 0.8)$ in metadata feature values, while those in italics were derived by inspecting the movies assigned to the clusters. }
\label{fig:western}
\end{figure}

\vskip 1em
\subsubsection{Case Study -- Westerns}
As a second case study, we looked at a set of eight clusters that were enriched at a level $\geq 0.8$ for a combination of two metadata feature values --  type ``feature film'' and genre ``western''. We constructed the subgraph of the normalised co-listed graph induced by the set of 227 movies in these clusters, which are linked by 1,234 edges. A visualisation of this graph is shown in \reffig{fig:western}. In general, the subgraph consists of a number of small weakly connected clusters of films from different eras, or with different leading actors.  

On the bottom right of \reffig{fig:western}, we see a large, densely-connected cluster that is highly enriched with movies associated with Italy. This group corresponds to so-called `Spaghetti Westerns', a darker sub-genre of Western films that emerged in the mid-1960s, with Italian directors or producers. The cluster in question contains a variety of films from well-known directors in this sub-genre, such as Sergio Leone (\eg \t{A Fistful of Dollars}, \t{Once Upon a Time in the West}) and Sergio Corbucci (\eg \t{Django}, \t{The Great Silence}). It is interesting to observe that these films are rarely assigned to the same lists as more traditional Westerns, such as those featuring John Wayne or Randolph Scott.

At the top of \reffig{fig:western}, we see a group of Westerns that do not immediately appear to share any other characteristics, beyond being produced in the United States. However, on inspecting the IMDb ratings scores for these movies, we see that they are frequently highly-rated -- the mean cluster rating is 7.4, 83\% of the movies have a score of 7.0 or higher. To contrast, the cluster of John Wayne films has a mean rating of 5.4. Examining individual list assignments, we see that the highly-rated cluster is comprised of movies assigned to lists such as ``Masterpieces of Western Genre'' and ``Western Writers of Americas 100 Greatest Westerns''. It is also interesting to note that two films (\t{Valdez is Coming}, \t{The Wild Bunch}) are assigned to this cluster, but also have edges connecting them to the Spaghetti Western cluster. Both films are recognised as being considerably influenced by the Italian Western movement, without actually belonging to that sub-genre. This is reflected in their IMDb list assignments and their resulting cluster memberships in our results.

\subsubsection{Case Study -- Feature Films from India}

\begin{figure}[!h]
\centering
\includegraphics[width=\linewidth]{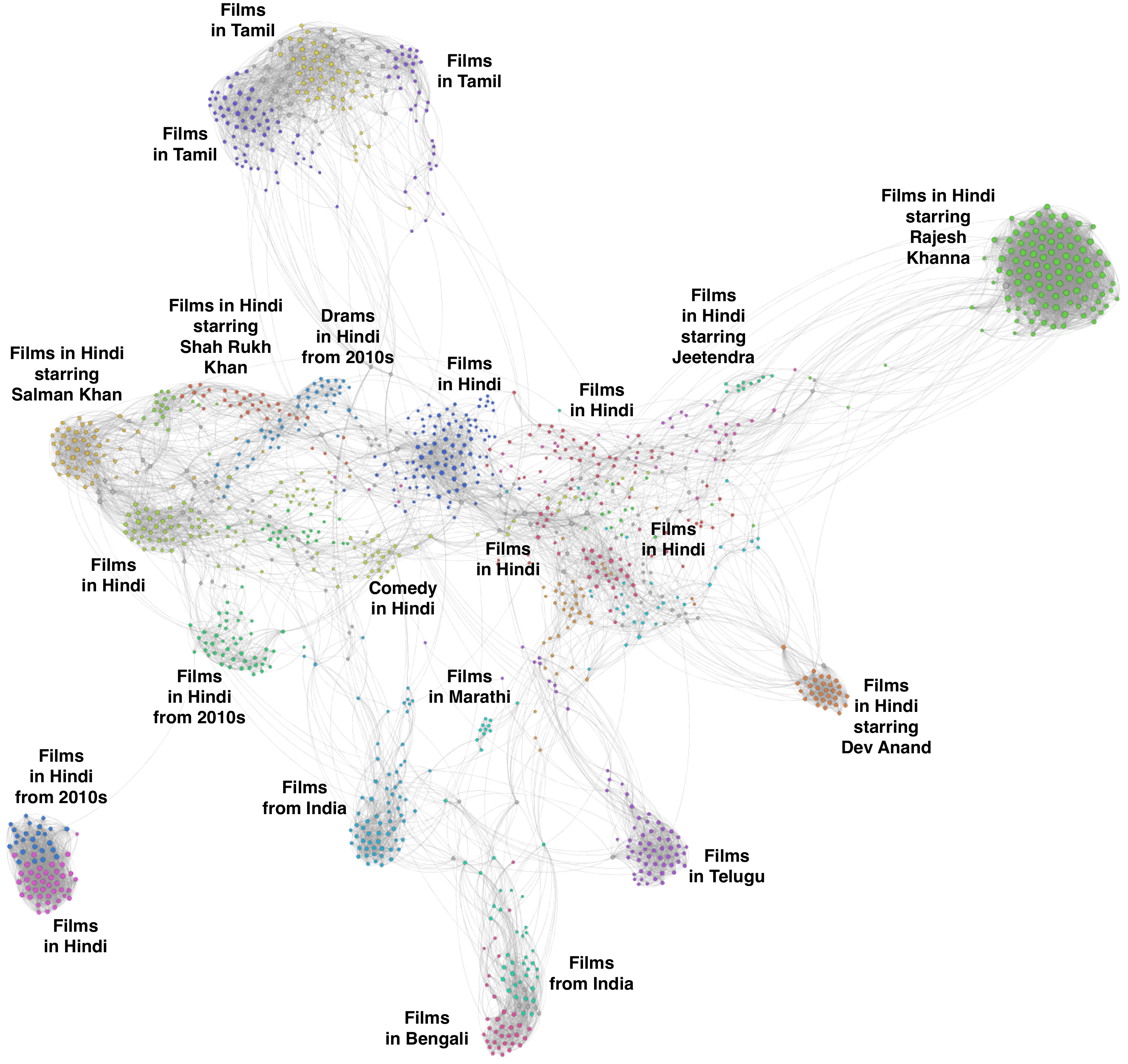}
\caption{A subgraph of the normalised co-listed graph, induced by a set of 30 clusters containing feature films associated with India. Nodes are coloured by cluster and scaled by degree. Nodes coloured in grey are assigned to multiple clusters. All cluster annotations were derived from high enrichment $(\geq 0.8)$ in metadata feature values.}
\label{fig:india}
\end{figure}

For our final case study, we examined the set of 30 clusters enriched at a level $\geq 0.8$ for type ``feature film'' and also India as country of origin. We constructed the corresponding subgraph of the normalised co-listed graph induced by the set of 1,359 movies in these clusters, which are linked by 14,703 edges. A force-directed layout of this subgraph is shown in \reffig{fig:india}. By inspecting the other metadata feature values aggregated on a per-cluster basis, we see that the majority of the clusters are also frequently highly-enriched with another feature -- \g{language}. Furthermore, inspecting the layout of the clusters in \reffig{fig:india} shows a clear division along linguistic lines. Perhaps unsurprisingly, films in Hindi are most prevalent in the subgraph, reflecting the broad accessibility of films in this language in Indian cinema and the fact that Hindi is the official language of the Union of India. We see a variety of clusters consisting of `Bollywood' (Hindi cinema) feature films. These are often further divided in clusters highly-enriched for films starring popular Bollywood actors, such as Shahrukh Khan (\eg \t{Kal Ho Naa Ho}), Dev Anand (\eg \t{Kalapani}), and Rajesh Khanna (\eg \t{Namak Haraam}). We also observe many smaller, more peripheral clusters that are enriched for films in other languages, such as Tamil, Telugu, and Bengali. It is interesting to observe that list curation activity on IMDb so clearly reflects the divide between Hindi Cinema and other Regional Cinemas.

\section{Conclusion}
\label{sec:conc}

In this paper, we propose the mining of valuable knowledge from the considerable efforts made by users to curate lists of items -- in our case, the mass creation of movie lists of users of the site \url{imdb.com}. We have looked at uncovering latent grouping of items, such as feature films and TV series, by constructing a suitable graph representation that encapsulates list co-assignments and then applying ensemble overlapping clustering to this graph. This approach has allowed us to identify highly-nuanced clusters of movies. In some cases these clusters reflect the personal tastes of IMDb users in a particular country, while in other cases the clusters reflect the evolution of movies of a particular type across several decades.

The graph representations and clusterings of movies discussed in this paper could be used in other contexts. 
For instance, the information derived from co-listed relations could have considerable relevance in movie recommender systems, which often rely on content-based features. 
While our focus has been on mining list information from IMDb, we suggest that the techniques proposed here have broader relevance in many other contexts, where the crowdsourced creation of lists is currently facilitated. This could range from the  mining Twitter lists to identify factions in a particular geopolitical region, to analysing collections of Foursquare lists in order to identify popular itineraries.

\section*{Acknowledgment}
\noindent This publication has emanated from research conducted with the financial support of Science Foundation Ireland (SFI) under Grant Number SFI/12/RC/2289.

\bibliographystyle{IEEEtran}
\bibliography{imdb-tr}

\begin{thebibliography}{10}
\providecommand{\url}[1]{#1}
\csname url@samestyle\endcsname
\providecommand{\newblock}{\relax}
\providecommand{\bibinfo}[2]{#2}
\providecommand{\BIBentrySTDinterwordspacing}{\spaceskip=0pt\relax}
\providecommand{\BIBentryALTinterwordstretchfactor}{4}
\providecommand{\BIBentryALTinterwordspacing}{\spaceskip=\fontdimen2\font plus
\BIBentryALTinterwordstretchfactor\fontdimen3\font minus
  \fontdimen4\font\relax}
\providecommand{\BIBforeignlanguage}[2]{{%
\expandafter\ifx\csname l@#1\endcsname\relax
\typeout{** WARNING: IEEEtran.bst: No hyphenation pattern has been}%
\typeout{** loaded for the language `#1'. Using the pattern for}%
\typeout{** the default language instead.}%
\else
\language=\csname l@#1\endcsname
\fi
#2}}
\providecommand{\BIBdecl}{\relax}
\BIBdecl

\bibitem{zhong13loves}
C.~Zhong, S.~Shah, K.~Sundaravadivelan, and N.~Sastry, ``Sharing the loves:
  Understanding the how and why of online content curation,'' in \emph{Proc.
  7th International AAAI Conference on Weblogs and Social Media}, 2013.

\bibitem{greene12recsys}
D.~{Greene}, G.~{Sheridan}, B.~{Smyth}, and P.~{Cunningham}, ``{Aggregating
  Content and Network Information to Curate Twitter User Lists},'' in
  \emph{Proc. 4th ACM RecSys Workshop on Recommender Systems \& The Social
  Web}, 2012.

\bibitem{imdbwiki}
Wikipedia, ``{Internet Movie Database},''
  \url{http://en.wikipedia.org/wiki/Internet_Movie_Database}, August 2013.

\bibitem{white81cocite}
H.~White and C.~Griffith, ``{Author Cocitation: A Literature Measure of
  Intellectual Structure},'' \emph{J. American Society for Information
  Science}, vol.~32, no.~3, pp. 163--171, 1981.

\bibitem{gmur03invisible}
M.~Gm\"{u}r, ``{Co-citation analysis and the search for invisible colleges: A
  methodological evaluation},'' \emph{Scientometrics}, vol.~57, no.~1, pp.
  27--57, 2003.

\bibitem{kim10lists}
D.~Kim, Y.~Jo, I.-C. Moon, and A.~Oh, ``Analysis of {Twitter} lists as a
  potential source for discovering latent characteristics of users,'' in
  \emph{Workshop on Microblogging at CHI'10}, 2010.

\bibitem{wu11says}
S.~Wu, J.~Hofman, W.~Mason, and D.~Watts, ``Who says what to whom on
  {Twitter},'' in \emph{Proc. 20th International Conference on World Wide
  Web}.\hskip 1em plus 0.5em minus 0.4em\relax ACM, 2011, pp. 705--714.

\bibitem{greene12ecml}
D.~Greene, D.~O'Callaghan, and P.~Cunningham, ``{Identifying Topical Twitter
  Communities via User List Aggregation},'' in \emph{Proc. 2nd International
  Workshop on Mining Communities and People Recommenders (COMMPER 2012) at ECML
  2012}, 2012, pp. 41--48.

\bibitem{garcia12lists}
A.~Garc\'{i}a-Silva, J.~Kang, K.~Lerman, and O.~Corcho, ``Characterising
  emergent semantics in {Twitter} lists,'' \emph{The Semantic Web: Research and
  Applications}, pp. 530--544, 2012.

\bibitem{basu98rec}
C.~Basu, H.~Hirsh, and W.~Cohen, ``Recommendation as classification: Using
  social and content-based information in recommendation,'' in \emph{Proc. 15th
  National Conference on Artificial Intelligence}, 1998, pp. 714--720.

\bibitem{herr07movies}
B.~W. Herr, W.~Ke, E.~Hardy, and K.~Borner, ``{Movies and Actors: Mapping the
  Internet Movie Database},'' in \emph{Proc. 11th International Conference on
  Information Visualization}.\hskip 1em plus 0.5em minus 0.4em\relax IEEE,
  2007, pp. 465--469.

\bibitem{dodds06bond}
K.~Dodds, ``{Popular geopolitics and audience dispositions: James Bond and the
  Internet Movie Database (IMDb)},'' \emph{Transactions of the Institute of
  British Geographers}, vol.~31, no.~2, pp. 116--130, 2006.

\bibitem{lanc12consensus}
A.~Lancichinetti and S.~Fortunato, ``Consensus clustering in complex
  networks,'' \emph{Sci. Rep.}, vol.~2, 03 2012.

\bibitem{hubert85compare}
L.~Hubert and P.~Arabie, ``Comparing partitions,'' \emph{Journal of
  Classification}, pp. 193--218, 1985.

\end{thebibliography}
\end{document}